\documentclass[5p,times,twocolumn]{elsarticle}
\usepackage[english]{babel}
\usepackage[utf8]{inputenc}
\usepackage[colorinlistoftodos, color=green!40, prependcaption]{todonotes}
\usepackage{amsthm}
\usepackage{mathtools}
\usepackage{physics}
\usepackage{xcolor}
\usepackage{graphicx}
\usepackage{placeins}
\usepackage[T1]{fontenc}
\usepackage{lipsum}
\usepackage{csquotes}
\usepackage[pdftex, pdftitle={Article}, pdfauthor={Author}]{hyperref} 
\begin{document}
\title{Scanning qubit probe of edge states in a topological insulator}

\author[add1]{Nicolas Delnour\corref{cor1}}
\ead{nicolas.delnour@mcgill.ca}
\author[add2]{Alexei Bissonnette}
\author[add3,add4]{Hichem Eleuch}
\author[add2]{Richard MacKenzie}
\author[add1]{Michael Hilke}
\address[add1]{Department of Physics, McGill University, Montreal, H3A 2T8, QC, Canada}
\address[add2]{Département de physique, Université de Montréal, Montréal, H3C 3J7, QC, Canada}
\address[add3]{Department of Applied Physics and Astronomy, University of Sharjah, United Arab Emirates}
\address[add4]{College of Arts and Sciences, Abu Dhabi University, Abu Dhabi, 59911, United Arab Emirates}
\cortext[cor1]{Corresponding author}


\date{\today} 

\begin{abstract}In this work, we propose a novel qubit-based sensor with the ability to characterize topological edge states in low-dimensional systems. A composite system is studied, consisting of a qubit coupled to a topologically nontrivial Su-Schrieffer-Heeger chain between semi-infinite lead channels. This qubit probe utilizes decoherence dynamics which, under a weak-coupling framework, are related to the environment’s local density of states. Qubit decoherence rate measurements along a sample therefore provide the means to extract edge state profiles. The environment’s influence on the qubit’s subspace  is captured by an effective projective treatment, leading to an analytical decoherence rate expression. We demonstrate that the scanning qubit probe identifies and yields a complete spatial characterization of the topological edge states within the composite system.

\end{abstract}

\maketitle

\section{Introduction} \label{sec:intro}

Efforts towards the miniaturization of technological components have naturally led to an increase in the research of low-dimensional materials. Due to their constrained dimensions, low-dimensional materials host interesting collective behaviours, some of which are linked to an interplay of topology and wavefunctions \cite{RevModPhys.88.035005}. The advent and rise of topological materials in the past couple of decades can be traced to peculiar exotic states of matter, such as topological edge states, that can exist within these materials \cite{RevModPhys.88.035005,Liu2019, shortcourse}. Signatures of nontrivial topologies in bulk materials are commonly studied by Angle-Resolved PhotoEmission Spectroscopy (ARPES), a powerful method to probe the momentum-space band structure of a material \cite{lv2015observation,xu2015discovery,lv2017observation,ma2018three}. A common signature of topological insulators is the gapless energy dispersion due to edge states and a nontrivial Berry phase. However, in one-dimensional systems, the dispersion is not sufficient and transport/transmission probes, which typically involve a study of the conductance through a sample, have been used. For instance, quasiparticle interference via the use of scanning tunnelling spectroscopy can be used to resolve backscattering-protected topological surface states \cite{roushan2009topological,yin2021probing}. Transmission probes can prove highly useful for studying edge states in lower-dimensional systems but cannot easily discern how the edge states are distributed along the boundaries of a system or whether a localized mode is on the left or right boundary of a one-dimensional chain \cite{ZAIMI2021127035}. Most other techniques used to probe topological materials including magnetotransport \cite{yan2017topological}, thermal conductance \cite{van2014thermal}, ultrafast dynamics \cite{marsi2018ultrafast}, and superconducting qubits for Majorana zero modes \cite{pekker2013proposal} cannot probe the spatial dependence of edge states.

This work exploits the intimate relationship between the decoherence of a qubit (two-level system) and its environment to provide a real-space  diagnostic of environment state amplitudes. To be concrete, rather than  striving to minimize decoherence, as is often the goal in quantum information technologies \cite{Fedorov:2004:1546-1955:132, BROADBENT20092489, 8914134, Burnett2019}, an investigation of how the environment affects the decoherence dynamics of a qubit reveals how these dynamics can be used, in reverse, to probe local properties of the qubit's surroundings. In effect, a novel quantum sensor is proposed, dubbed the decoherence probe, which utilizes as its mechanism of action the natural measurement induced from interactions with an environment to offer superior position-basis diagnostics for low-dimensional topological systems. As opposed to probes that are sensitive to the dispersion, such as ARPES \cite{lv2019angle} and our earlier work on a localized qubit \cite{ZAIMI2021127035}, the decoherence probe presented in this work accesses the entirety of real-space and can be utilised to map out the spatial amplitude of populated states at a given energy. 

This probe has potential for applications to low-dimensional engineered topological systems and  various mesoscopic systems. For definiteness, we demonstrate the decoherence probe’s ability to study the topological edge states arising in the Su-Schrieffer-Heeger (SSH) model. The SSH model, owing to its one-dimensional nature, is widely considered the simplest model of a topological material. The model's simplicity has made it a promising platform for the design of engineered/synthetic topological materials \cite{PhysRevX.11.011015, Kiczynski2022, PSJ2022, St_Jean_2017,PhysRevLett.126.043602,PhysRevLett.128.203602,topodissip2022,PhysRevLett.124.023603}. This work demonstrates the probe's complete spatial characterization capabilities. The full system to be probed is composed of a finite SSH chain between two semi-infinite conducting leads. Semi-infinite leads model an extended environment and allow for the decoherence of the qubit. Furthermore, two semi-infinite leads are chosen as a \textit{realistic} constraint to model the response of  systems where the topological material of interest is integrated within some circuit geometry \cite{Kiczynski2022, PhysRevX.11.011015}. The qubit decoherence rate is shown to provide a measure of the local density of states (LDOS) within the SSH chain, local to where the qubit was coupled. Weak coupling is considered both for the SSH-qubit and SSH-lead coupling. The former ensures an analytically tractable expression for the qubit's  decay, while the latter ensures that the eigenstates of the coupled SSH chain differ from the isolated chain only up to some perturbative correction. Various open system configurations of the SSH model, including open boundary couplings such as leads or $\mathcal{PT}$-symmetric defect terms, have been previously studied in literature \cite{OSTAHIE2021127030,PhysRevB.95.115443,PhysRevA.98.013628,PhysRevA.89.062102, PhysRevLett.127.250402}.

This work is structured as follows. The composite system consisting of the SSH chain, semi-infinite leads, and qubit is introduced and detailed  in Section \ref{sec:composite}. Topological edge states and their properties are briefly reviewed, followed by the detailed treatment of the qubit and the extraction of its decoherence rate. In Section \ref{sec:deco}, the composite system is studied using the decoherence of the qubit in order to obtain a full spatial characterization of SSH edge states for chains comprising both an even and odd number of sites.

\section{The Composite System} \label{sec:composite}
\begin{figure}[ht]
\centering
\includegraphics[width=0.9\linewidth]{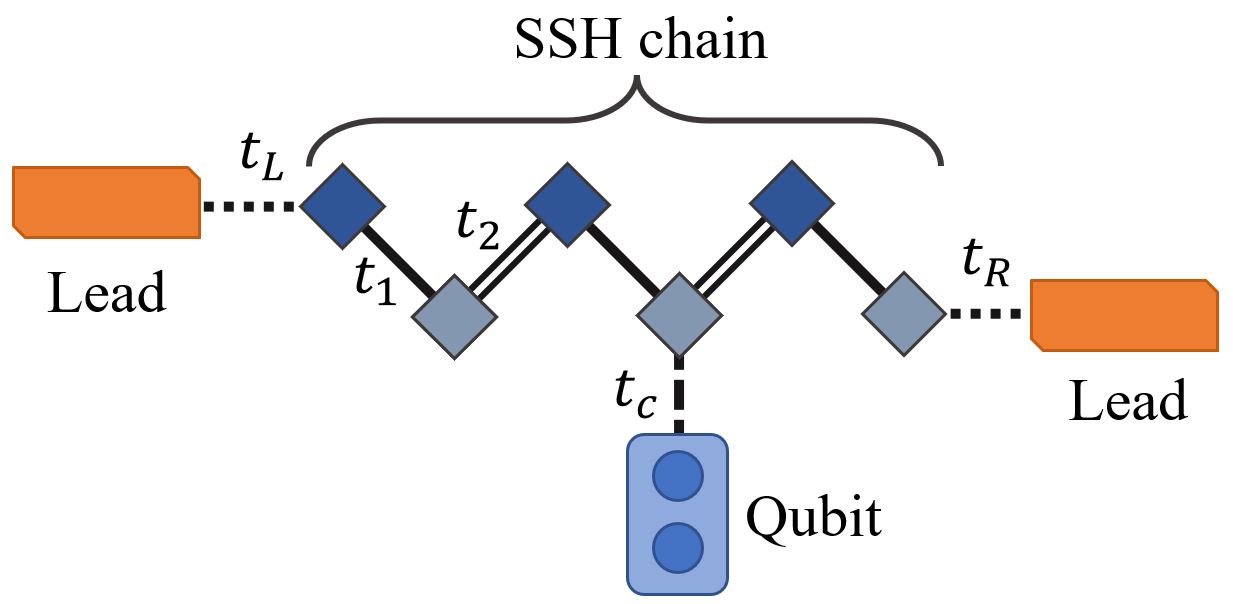}%
\caption{Composite system geometry for the qubit probe shown here for an SSH chain of length $N=6$.}\label{tripartgraph}
\end{figure}
Consider the mesoscopic system illustrated in Fig.~\ref{tripartgraph} consisting of a finite SSH chain of length $N$, two semi-infinite leads, and a qubit (which can be viewed more generally as a two-level system). 

The SSH chain is described by the Hamiltonian 
\begin{align}\label{HSSH}
H_{\text{SSH}} = & \, \, \, \sum_{m=1}^{N-1} t_m \big( c^\dagger_m c_{m+1} + c_{m+1}^\dagger c_m\big),
\end{align}
where $m$ is the site index, $c_m^\dagger,\ c_m$ are the creation and annihilation operators for site $m$, and the hopping parameter $t_m$ between sites $m$ and $m+1$ alternates between two real values, $t_1$ for odd $m$ and $t_2$ for even $m$.

The left and right semi-infinite leads are described by the Hamiltonians
\begin{equation}\label{Hsemiinf}
\begin{split}
H_{L,\infty} = & \sum_{m=1}^{\infty} t \big( l_m^\dagger l_{m+1} +  l_{m+1}^\dagger l_{m} \big),\\
H_{R,\infty} = & \sum_{m=1}^{\infty} t \big( r_m^\dagger r_{m+1} +  r_{m+1}^\dagger r_{m} \big),
\end{split}
\end{equation}
where again $m$ is the site index and where $l$, $r$ are the respective site creation and annihilation operators for the two leads.

The double dot qubit is described by the Hamiltonian
\begin{equation}\label{HDD}
H_{\mathrm{DD}}=  \epsilon_{1}d_1^\dagger d_1 +\epsilon_{2}d_2^\dagger d_2 +\tau \left( d_1^\dagger d_2 + d_2^\dagger d_1 \right) ,
\end{equation}
where $\epsilon_1$, $\epsilon_2$ and $\tau$ are supposed real and $d$ represent creation and annihilation operators for the two sites. 

The subsystems are coupled together as displayed in Fig.~\ref{tripartgraph}. In what follows, the normalization of all the parameters in the Hamiltonian will be chosen such that the lead hopping parameter is set to $t=1$. The left (right) lead is coupled to the first (last) SSH site with coupling $t_L$ ($t_R$), while the second qubit site is coupled to the $n^{\mathrm{th}}$ SSH site with coupling $t_c$.
The leads serve as the extended environment through which the qubit ultimately decoheres. For simplicity, we will later set $t_L=t_R$, ensuring a symmetric qubit  response over the SSH chain. Putting everything together, the full Hamiltonian for the system is
\begin{equation}\label{eq1}
\begin{split}
H  &= \, H_{\mathrm{DD}}+H_{\text{SSH}}+H_{L\infty}+H_{R\infty} \\&
+t_c \big(c_n^\dagger d_2 + d_2^\dagger c_n\big)+t_L \big( c_1^\dagger l_1 + l_1^\dagger c_1\big)\\&
+t_R \big( c_N^\dagger r_1 + r_1^\dagger c_N\big).
\end{split}
\end{equation}

\subsection{Topological Edge States of the SSH Model}
It is useful to highlight the spectral and real-space properties of the model's topological edge states, as they are the central feature chosen to demonstrate the probe's spatial characterization abilities. Analytical solutions to the model along with energy spectra are obtained and detailed in previous work  \cite{ZAIMI2021127035}. The distinct phases within the SSH model are defined by a topological invariant, the winding number $\nu$, which can be computed from the closed trajectory of the bulk Hamiltonian vector $\Vec{h}(k)=(t_1+t_2 \cos{2k},\, t_2 \sin{2k}, \,0)$, expressed here in the Pauli matrix basis, sweeping through the Brillouin Zone (BZ) \cite{shortcourse}. Note that the BZ is defined here for the wavevector range $k \in (-\pi/2, \pi/2$]. The winding number is intimately tied to the staggered nature of the hopping terms and can be related to the parameter $r=t_1/t_2$. 
\begin{figure}[ht]
\centering
\includegraphics[width=1\linewidth]{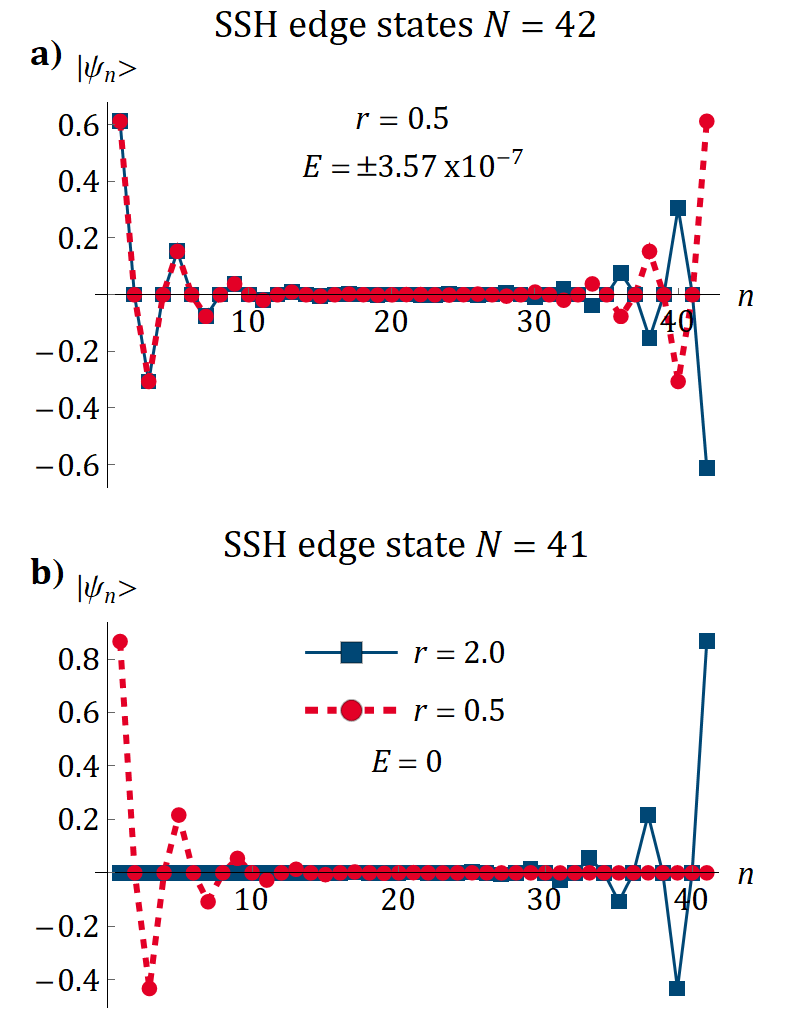}%
\caption{ \textbf{a)} Pair of topological edge states arising for $r<r_C$ in a finite SSH chain of even-length. \textbf{b)} Edge state for a finite SSH chain of odd-length. A single edge state is always present in the odd-length SSH chain, with localization occurring on the left (right) boundary for $r<1$ ($r>1$).}\label{sshedgegraph}
\end{figure}

For finite even-length SSH chains with $N=2M$ (where $M$ denotes the number of unit cell), the winding number is $\nu=0$ when $r>r_C=N/(N+2)$, where there are $N$ bulk states and the chain is said to be in the trivial phase. The critical parameter $r_C$ as obtained in \cite{ZAIMI2021127035,delplace_zak_2011} is a feature of finite chains and  reproduces the expected topological phase transition at $r_C=1$ in the thermodynamic limit, $N \rightarrow \infty$. When $r<r_C$, the winding number becomes $\nu=1$ and the SSH chain is in a topological phase. In this configuration, there exist $N-2$ bulk states and the two remaining solutions take the form of near-zero-energy modes. From the dispersion relation, $E(k)= \pm \sqrt{t_{1}^{2}+t_{2}^{2}+2 t_{1} t_{2} \cos{2k}}$, states lying within the gap must have complex wavenumbers $k \rightarrow \pi/2 + i\kappa$  and, hence, are excitations localized at the boundaries of the open chain, forming so-called edge states. In the thermodynamic limit, edge states tend towards zero energy and get pinned at the midgap. This differs to open finite chains, where a splitting about the Fermi level gives rise to small but non-zero energies for these modes. As a direct consequence of chiral symmetry, a non-zero-energy eigenstate has equal support on both sublattices, contrary to the $E=0$ eigenstates which are confined to a single sublattice. As the edge state energies tend exponentially to zero with $N\rightarrow \infty$, one observes an apparent confinement near the boundaries of the finite chain, but  with hybridization on both odd and even sites taking place  in the bulk, as seen in  Fig.~\ref{sshedgegraph}a. 

An important difference for finite SSH chains with an odd number of sites, $N=2M+1$, is that there always exist $N-1$ bulk solutions and a single localized mode pinned to the midgap regardless of the value of $r$. Chiral symmetry, which imposes symmetry of the spectrum, forces the unpartnered eigenvalue to $E=0$ and leads to the perfect confinement of the edge state on a single sublattice of the chain, as shown in Fig.~\ref{sshedgegraph}b for two values of $r$.

\subsection{Effective treatment of the Qubit}
We refer the reader to \cite{ZAIMI2021127035} for a more thorough discussion of the qubit described by (\ref{HDD}). The time evolution of level transitions is obtained from the off-diagonal element of the Green's function of the double dot qubit: $\left|G^{12}_{\mathrm{DD}}(t)\right|^{2}$, where $G_{\mathrm{DD}}(E)=(E-H_{\mathrm{DD}})^{-1}$. The eigenvalues and energy splitting of the isolated qubit are given by 
\begin{equation}
\lambda_{\pm}=\epsilon_{0} \pm \frac{\delta}{2} \quad , \quad \delta=\sqrt{\delta_{0}^{2}+4 \tau^{2}},
\end{equation}
where, $\epsilon_0 =(\epsilon_1 +\epsilon_2)/2$ is the average energy of the two basis states and $\delta_0=(\epsilon_1 - \epsilon_2)$ is the energy splitting of the uncoupled basis states.  With the goal of characterizing the time evolution of the qubit embedded within an environment, an effective projective approach, detailed in \ref{appendix1} for an arbitrary environment, is used. This description captures the influence of the environment through a self-energy term added to $G_{\mathrm{DD}}$ in a new Green's function $G'_{\mathrm{DD}}$, to which one can associate the effective Hamiltonian $H'_{\mathrm{DD}}$ satisfying $G'_{\mathrm{DD}}=(E-H'_{\mathrm{DD}})^{-1}$. This technique is exact within the $2\times2$ subspace of the qubit, and is used to treat the qubit's response within the composite system, as will now be detailed.

Starting from (\ref{eq1}) and following the method in  \ref{appendix1}, the SSH-lead blocks are addressed first and reduced to an $N\times N$ effective Hamiltonian $H_{\mathrm{SSH},\infty}$ for the subspace of the SSH chain reflecting the influence of the leads,
\begin{equation}\label{sshleadscombined}
     H_{\mathrm{SSH}, \infty} \equiv H_{\text{SSH}} + \Sigma_L c_1^\dagger c_1 + \Sigma_R c_N^\dagger c_N.
\end{equation}
This is exactly $H_{\mathrm{SSH}}$ with the addition of self-energies taking the form of complex on-site potentials at the boundary elements of the SSH chain. The self-energies $\Sigma_{L,R}$ originate from, and account for, the left and right leads, respectively. As the leads are surface-coupled to the SSH chain, 
\begin{equation}\label{GLEAD}
\Sigma_{L,R}(E)=t_{L,R}^2 G_{\infty}^S(E),
\end{equation}
where
\begin{equation}\label{GLEAD1}
G_{\infty}^S(E) = \frac{1}{2} \left(E -\mathrm{sgn}(E+2)\sqrt{E^2-4}\right)
\end{equation}
is the surface term of the semi-infinite lead's Green's function \cite{ZAIMI2021127035,datta_2005,economou}. The effective Hamiltonian for the $(N+2)\times(N+2)$ subspace of the SSH chain and qubit is now 
\begin{equation}\label{2by2HSSHinfDD}
    H = H_{\text{DD}} + H_{\mathrm{SSH},\infty} +t_c \big(c_n^\dagger d_2 + d_2^\dagger c_n\big),
\end{equation}
and is further reduced to the $2\times 2$ effective Hamiltonian for the qubit subspace:
\begin{equation}\label{effectivequbit}
H'_{\mathrm{DD}}= H_{\mathrm{DD}} +  \Sigma^{nn}_{\mathrm{SSH},\infty}d_2^\dagger d_2,
\end{equation}
where the environment's self-energy $\Sigma^{nn}_{\mathrm{SSH,\infty}}(E)=t^2_c G^{nn}_{\mathrm{SSH,\infty}}(E)$ stems from the combined SSH-lead environment to which the qubit is coupled at a site $n$ with strength $t_c$. The superscript $G^{nn}$ denotes the $(n,\,n)$ diagonal element of $G$. This is an important distinction compared to previous work \cite{ZAIMI2021127035}, where only the surface coupling between the qubit and the end of a chain was considered. The derivation of the diagonal elements of the Green's functions satisfying $G_{\mathrm{SSH},\infty}(E) = (E-H_{\mathrm{SSH},\infty})^{-1}$ is detailed in   \ref{gnnsshinfderiv}. The four analytical forms for $G^{nn}_{\mathrm{SSH},\infty}(E)$, which depend on the parity of the SSH chain length $N$ and the parity of the sublattice site $n$, are given below. We find, for chains of odd length $N$,
\begin{equation}\label{Godd}
G^{nn}_{\mathrm{SSH},\infty}(E)= \begin{cases} \big[ E + g_{L1} +h_{R1}(t_1,t_2)  \big] ^{-1} & \text{for $n$ odd}\\
\big[ E + g_{L2} +h_{R2}(t_1,t_2) \big] ^{-1} &\text{for $n$ even} \end{cases},
\end{equation}
while for SSH chains of even length $N$,
\begin{equation}\label{Geven}
G^{nn}_{\mathrm{SSH},\infty}(E)= \begin{cases} \big[ E + g_{L1} +h_{R2}(t_2,t_1)  \big] ^{-1} & \text{for $n$ odd}\\
\big[ E + g_{L2} +h_{R1}(t_2,t_1) \big] ^{-1} &\text{for $n$ even} \end{cases}.
\end{equation}
The terms in $G^{nn}_{\mathrm{SSH},\infty}(E)$ are given below,  adopting the shorthand notation $\sin{(Ak)}=s_A$:
\begin{align}\label{gcoeff}
g_{L1}=&\frac{t_{2}^{2} E s_{n-1}-t_{2} \Sigma_{L}\left(t_{1} s_{n-3}+t_{2} s_{n-1}\right)}{E \Sigma_{L} s_{n-1}-t_{2}\left(t_{1} s_{n+1}+t_{2} s_{n-1}\right)}, \\[10pt]
g_{L2}=&\frac{t_{1} \Sigma_L E s_{n-2}-t_1 t_{2} \left(t_{1} s_{n}+t_{2} s_{n-2}\right)}{E t_2 s_{n}-\Sigma_L \left(t_{1} s_{n-2}+t_{2} s_{n}\right)},\\[10pt]
\begin{split}
&\hspace{-6.5cm} h_{R1}(t_1, t_2) = \cr \hspace{-1cm}\frac{t_{1}^{2} E s_{N-n}-t_{1} \Sigma_{R}\left(t_{1} s_{N-n}+t_{2} s_{N-n-2}\right)}{E \Sigma_{R} s_{N-n}-t_{1}\left(t_{1} s_{N-n}+t_{2} s_{N-n+2}\right)},
\end{split}\\[10pt]
\begin{split}
&\hspace{-7.5cm}h_{R2}(t_1, t_2) =\cr \hspace{-1cm} \frac{t_{2} E \Sigma_R s_{N-n-1}-t_{1} t_2 \left(t_{1} s_{N-n-1}+t_{2} s_{N-n+1}\right)}{E t_1 s_{N-n+1}-\Sigma_R \left(t_{1} s_{N-n+1}+t_{2} s_{N-n-1}\right)}.
\end{split}
\end{align}
Note that the expressions (\ref{Godd},\ref{Geven}) are arranged such that the second terms stem from the interface between the left lead and the SSH chain, while the third terms correspond to the right lead/chain-interface. 

Solving for the eigenvalues of (\ref{effectivequbit}), one finds
\begin{align}\label{lambdaprimE}
\lambda'_{\pm}(E)&=\frac{1}{2}\left(\epsilon_{1}+\epsilon_{2} +\Sigma^{nn}_{\mathrm{SSH,\infty}}(E) \pm \delta'(E)\right)\nonumber\\
&=\epsilon_{0} + \frac{\Sigma^{nn}_{\mathrm{SSH,\infty}}(E)\pm\delta'(E)}{2} ,
\end{align}
where $\delta'(E)$ is the energy splitting of the open qubit:
\begin{equation}\label{deltaprim}
\delta'(E)=\sqrt{(\delta_{0}-\Sigma^{nn}_{\mathrm{SSH,\infty}}(E))^{2}+4 \tau^{2}}.
\end{equation}
Unsurprisingly, the above expressions reduce to the isolated qubit case when $\Sigma^{nn}_{\mathrm{SSH,\infty}}(E) = 0$.

\subsection{Extracting and Interpreting the Decoherence  Rate}\label{decointerp}
The Green's function associated with  (\ref{effectivequbit}) is
\begin{equation}\label{greensDD}
G_{\mathrm{DD}}'(E)=\left(\begin{array}{cc} E-\epsilon_1 & -\tau \vspace{0.1cm} \\ -\tau  \qquad & E-\epsilon_2 -\Sigma^{nn}_{\mathrm{SSH,\infty}}(E)  \end{array}\right)^{-1},
\end{equation}
given here in matrix notation following (\ref{appendixGeff}). The decay of the qubit can again be linked to the off-diagonal elements of the time-dependent Green's function. Unlike the isolated qubit, however, the energy dependence of $\Sigma^{nn}_{\mathrm{SSH,\infty}}(E)$ in $G'_{\mathrm{DD}}(E)$ precludes a direct analytical evaluation of the Fourier transform for $G'_{\mathrm{DD}}(t)$. Qualitatively, the effective description is non-Hermitian and the self-energies lead to complex eigenvalues such that $G'_{\mathrm{DD}}(t)$ is no longer purely oscillatory and instead takes the form of a damped exponential, reflecting the decay of the qubit due to the open nature of the system. A perturbative weak-coupling approximation with $t_c^2 N \ll 1$ is used to obtain an analytical form for the time-dependent Green's function as detailed in Appendix A of \cite{PhysRevA.95.062114}.  The decay rate expression extracted from $G'_{\mathrm{DD}}(t)$, taking into account a weak coupling to an arbitrary site $n$, yields 
\begin{equation}\label{tauinv}
    (\tau_\phi^{nn})^{-1} \approx \mathrm{min} \left(-\frac{1}{2}\mathrm{Im} \left\{ \Sigma^{nn}_\mathrm{SSH,\infty}(\lambda_\pm) \pm \delta'(\lambda_\pm) \right\} \right).
\end{equation}
Note that, as a result of the perturbative treatment, the energy-dependent quantities in $(\tau_\phi^{nn})^{-1}$ are evaluated at the eigenvalues of the isolated qubit, $\lambda_\pm$. The choice of $\lambda_+$ or $\lambda_-$ is that which gives the slowest decay, since this term dominates the long-time decoherence of the qubit.

It is useful to reflect on (\ref{tauinv}) to understand what physical quantity can be probed through a study of decoherence rates. Evidently, the first term in (\ref{tauinv}) is proportional to the negative imaginary part of  $G^{nn}_{\mathrm{SSH},\infty}(E)$. This is precisely the definition of the local density of states (LDOS) at a site $n$ within the SSH chain,
\begin{equation}
    d(E,n)=-\mathrm{Im}\{G^{nn}_{\mathrm{SSH},\infty}(E)\}.
\end{equation} The term $\mathrm{-Im}\{\Sigma^{nn}_{\mathrm{SSH},\infty}(\lambda_\pm) \}$ in \eqref{tauinv} is therefore directly proportional to the LDOS in the SSH chain at the energies of the isolated qubit. The second term, $ \delta'(\lambda_\pm)$, is more complex. There is a dependence to the LDOS, but a linear proportionality between $\mathrm{Im} \{ \delta' \}$ and $\mathrm{Im}\{\Sigma^{nn}_{\mathrm{SSH},\infty}\}$  depends on the relationship between the self-energy and the qubit parameters. Factoring $\Sigma^{nn}_{\mathrm{SSH},\infty}$ and expanding the root in (\ref{deltaprim}) for  $\Sigma^{nn}_{\mathrm{SSH},\infty} \gg (\delta_0$, $\tau )$ yields
\begin{align}\label{simped}
    \delta'&=\Sigma^{nn}_{\mathrm{SSH},\infty} \bigg( 1 - \frac{\delta_0}{ \Sigma^{nn}_{\mathrm{SSH},\infty}} + \mathcal{O}\left(\frac{1}{\Sigma^{nn\quad 2}_{\mathrm{SSH},\infty}}\right) + ... \bigg) \nonumber\\
    &\approx \Sigma^{nn}_{\mathrm{SSH},\infty},
\end{align}
with small corrections of order $\delta_0$. We can then write to a good approximation that $ - \mathrm{Im} \{ \delta'\}  \approx - \mathrm{Im} \{ \Sigma^{nn}_{\mathrm{SSH},\infty}\}$. Note that (\ref{simped}) leads to a nearly perfect cancellation for the appropriate upper or lower sign in (\ref{tauinv}) for a given  $\delta'(\lambda_\pm)$, showing that this near-cancellation is crucial in extracting the slower of the two competing decay rates. Choosing TLS parameters $\delta_0$ and $\tau$ small with respect to $\Sigma^{nn}_{\mathrm{SSH},\infty}$ we obtain, for the appropriate $\lambda_+$ or $\lambda_-$ which minimizes the decay rate, an interesting conclusion:
\begin{align}\label{tauDOS}
 (\tau_\phi^{nn})^{-1} & \propto -\mathrm{Im} \{ G^{nn}_{\mathrm{SSH},\infty}(\lambda_\pm)\} = d(\lambda_\pm,\,n) .
\end{align}
This useful interpretation of the decoherence rate as proportional to the LDOS is conceptually reasonable; couplings to densely populated sectors of the environment's energy spectrum should lead to faster decoherence than couplings to an empty or sparse sector.

\section{Scanning Decoherence Probe} \label{sec:deco}

The decoherence rate (\ref{tauinv}) depends on the self-energy $\Sigma_{\mathrm{SSH},\infty}^{nn}(\lambda_\pm)$ and so is a function of both the eigenvalues $\lambda_\pm$ of the qubit and the coupling site $n$ between the qubit and the SSH chain. This, paired with the proportionality between $(\tau_\phi^{nn})^{-1}$ and the LDOS, can be used to extract the amplitude profiles of states residing within the SSH chain.  Tuning one of the qubit frequencies to the midgap effectively couples the qubit to edge states, and decoherence rates can be used to obtain spatial characteristics of these states. 

The probe can be implemented in two ways. The first is a scanning probe approach where the qubit is coupled to a site $n$ and initialized such that it possesses an eigenvalue, say $\lambda_-$, tuned to an energy of interest in the SSH chain. The decoherence rate of the qubit is measured as it naturally decoheres due to its environment coupling. This process can be repeated at all sites along the chain, reinitializing the qubit at $\lambda_-$ every time, and yields a spatial profile of states in the SSH chain at energy $\lambda_-$. Alternatively, multiple qubits can be integrated along the entire length of a mesoscopic system as done in \cite{PhysRevX.11.011015} to provide simultaneous readouts along the sample of interest. LDOS profiles are therefore extracted from the decoherence rate as a function of the site number. Additionally, the DOS can be obtained by simply summing the LDOS over all sites. It has been demonstrated in previous work that the decoherence probe coupled to the end of the SSH chain is sensitive to the presence or absence of edge states \cite{ZAIMI2021127035}. We now show that scanning the probe along the length of the SSH chain can map out the profile of edge states. Results for both SSH chains of even and odd length are presented. 

\subsection*{Spatial characterization of topological edge states}

With the goal of first studying the topological edge states of an even-$N$ SSH chain, the qubit is initialized such that it has an  eigenvalue $\lambda_-$ tuned in the gap at the edge state energy. Since $\mathrm{Im}\left\{G^{nn}_{\mathrm{SSH},\infty}(E)\right\}$ is symmetric, we need not worry about the sign of the edge state energy we input; chiral partner states of equal and opposite energy have  identical amplitudes $| \psi_{\mathrm{edge}}|^2$ and will yield identical decoherence rate measurements. Recall that only the slowest decay rate is considered relevant in the treatment of (\ref{tauinv}); fine-tuning of the qubit parameters may be necessary to ensure that the desired $\lambda_-$ contribution remains minimal at all sites $n$ of the SSH chain. This allows one to precisely map the edge state over the entire length of the chain. Notably, qubit  parameters should be chosen such that $\lambda_+$ is resonant with a bulk state. In this way, bulk contribution $\lambda_+$ to the decay rate is  maximal and the decoherence rate actually reflects the $\lambda_-$ contribution -- the coupling to an edge state as desired. 
\begin{figure}[ht]
\begin{center}
{\includegraphics[width=1\linewidth]{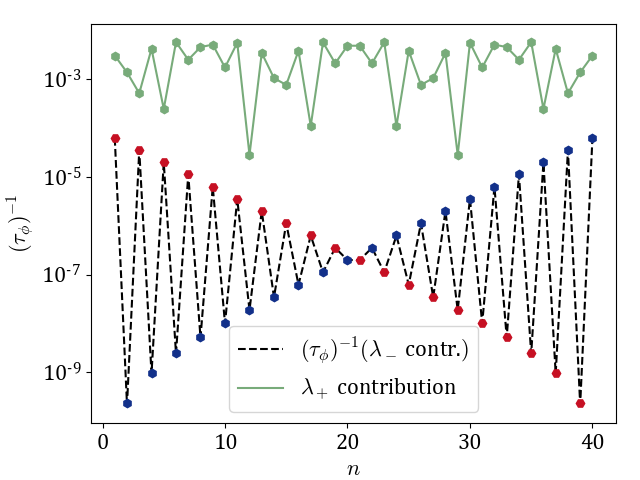}}
\end{center}
\caption{Plot of the decoherence rate (dashed black line) and its two  $\lambda_\pm$ contributions for a scan of a topological SSH chain of length $N=40$ with $t_1=0.6$ and $t_2=0.8$. Qubit parameters are tuned to ($\epsilon_1$, $\epsilon_2$, $\tau$) = (0.4017, 0.0034, 0.03) such that $\lambda_+=0.4039$ is tuned to a densely populated sector of the bulk bands while the $\lambda_-=1.11\times10^{-3}$  contribution, tuned to the edge state energy (red odd sites, blue even sites), is minimal all along the SSH chain. Coupling parameters are $t_c = 0.003$ and $t_L= t_R = 0.02$. Edge states have the largest penetration into the bulk for $r$ less than but close to $r_C$. Note the beyond-exponential edge state suppression expected in near-zero energy states of finite even-length SSH chains, accentuated here from choosing $r = 0.75$, relatively near $r_C=0.95$.}\label{NevenProbe}
\end{figure}

A plot of the decoherence rate $(\tau^{nn}_\phi)^{-1}$ as a function of $n$ for a topological SSH chain with $N=40$ and $r=0.75$ ($r<r_C$) is shown in Fig.~\ref{NevenProbe}. Different $\lambda_\pm$ contributions to the decoherence rate are shown, with $(\tau^{nn}_\phi)^{-1}$ being the minimum of the two. The decoherence rate is further decomposed into even and odd sites for visual clarity; see the figure caption for details. We note a stark contrast in the behaviour of both decay contributions. This is the result of $\lambda_+=0.4039$ residing within the bulk bands, which span the absolute energy range $\{0.1909,\,2.009\}$, while $\lambda_-=1.11\times10^{-3}$ is tuned in the gap to the energy of the topological edge state of the finite SSH chain. Thus, on the one hand, the $\lambda_+$ terms are  associated with an oscillatory bulk LDOS. On the other hand, the $\lambda_-$ terms, corresponding here to $(\tau^{nn}_\phi)^{-1}$, show quasi-exponential behaviour as a function of $n$. Near the left (right) edge, decoherence is rapid on odd (even) sites and gets suppressed as one moves into the bulk, reflecting the localization of the probed excitations. Additionally, even- (odd-) site decoherence is heavily suppressed in the vicinity of the left (right) edge, a result of edge states having near-zero amplitudes on these sites. Strong oscillations between the $\lambda_-$ contributions on even and odd sites when scanning along the chain are observed near zero energy.  These behaviours are in agreement with open SSH chain topological edge states which give an impression of sublattice confinement near the boundaries. 

\begin{figure}[ht]
\begin{center}
{\includegraphics[width=1\linewidth]{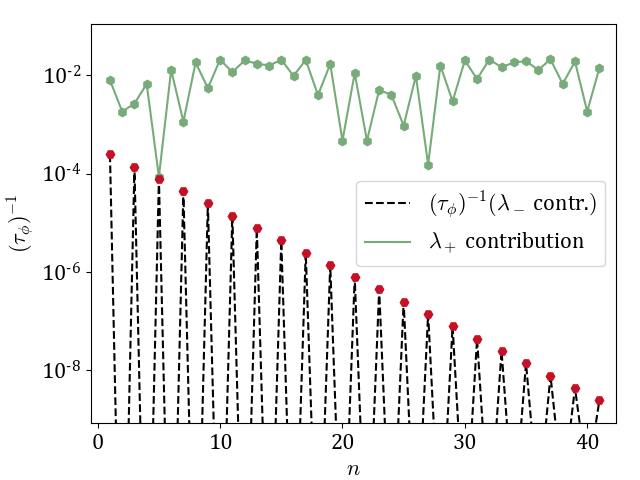}}
\end{center}
\caption{Both $\lambda_\pm$ contributions to the qubit decoherence rate as a function of position along an SSH chain of length $N=41$ are shown for $r=0.75$. The decoherence rates on odd sites are highlighted in red. The qubit is initialized with parameters ($\epsilon_1$, $\epsilon_2$, $\tau$) = (0.2843, 0.0031, 0.03), yielding $\lambda_-=0$ and the bulk-resonant frequency $\lambda_+ = 0.2875$ such that the zero energy contribution to the decay is always minimal. Coupling parameters are $t_c = 0.003$ and $t_L= t_R = 0.02$. The decoherence rate is indicative of an SSH chain with a left edge state: as expected of the zero-energy states, the decay rate shows an exponentially decaying profile entering the bulk.}\label{Noddprobe}
\end{figure}

The decoherence probe is now used to study odd-$N$ SSH chains and their respective edge states. In such chains, there always exists a zero-energy mode with localization on the left (right) edge for $r<1$ ($r>1$) as was shown in Fig.~\ref{sshedgegraph}b. As a  result of being pinned to the midgap, these states are confined to the odd-site sublattice with a pure exponential fall-off in the bulk. The qubit is tuned to have to $\lambda_-=0$ to probe the edge state. A decoherence plot for the $r<1$ phase for a chain of length $N=41$ is shown in Fig.~\ref{Noddprobe}. The case for the $r>1$ configuration is symmetric to Fig.~\ref{Noddprobe} under spatial inversion. The $\lambda_+$ contribution to the decoherence rate is again oscillatory from coupling to  bulk states. Looking at the $\lambda_-=0$ terms (odd sites highlighted in red), one observes large decoherence rates on the left edge and exponentially suppressed decoherence rates as one travels through the bulk on odd sites, indicative of a zero-energy state localized at the left edge. Strong $(\tau^{nn}_\phi)^{-1}$ oscillations arise on odd and even sites, with even-site decoherence rates being many orders of magnitude smaller than on odd sites. In principle, these edge states have no support on even sites, and the non-zero decay rates on these sites can be attributed to hybridization with lead states at the midgap, which is kept negligible by choosing $t_{L,R}\ll 1$. In the absence of fine tuning, which guarantees that the decay rate only corresponds to $\lambda_-$ terms, the probe can be insensitive to the complete edge state profiles and may instead sample bulk-oscillatory behaviour. The site $n=5$ in Fig.~\ref{Noddprobe} is an example of the bulk LDOS competing strongly with the edge state LDOS. Edge state signatures could  nonetheless be probed by seeking out exponential-like suppression of the decay rate deeper in the chain, where the lower amplitude of the edge states yields smaller decoherence rates. 
\begin{figure}[ht]
\begin{center}
{\includegraphics[width=1\linewidth]{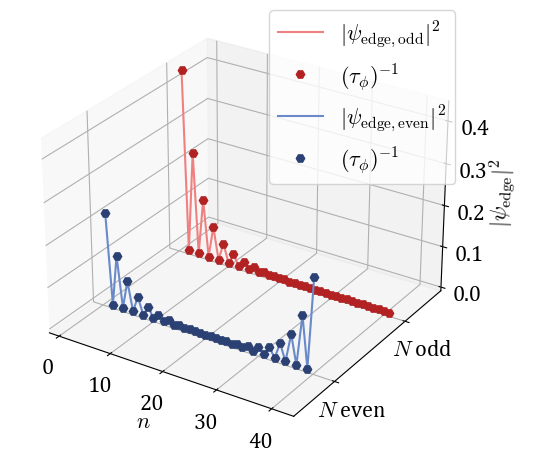}}
\end{center}
\caption{Comparison of the normalized LDOS profile extracted from the decoherence rate to the edge state profile for an open SSH chain of length $N=40$ (blue) and $N=41$ (red), both with $r = 0.75$. The LDOS is normalized to match the edge state amplitude. }\label{profileES}
\end{figure}
The above analysis demonstrates the sensitivity of the scanning decoherence probe to various edge states that arise in the SSH chain within the composite system. This includes the ability to differentiate edges states from even-$N$ and odd-$N$ SSH chains as well as the ability to know which boundary the state is localized on in the case of odd-$N$ zero-modes. Through its decay rate proportionality to the LDOS, the  probe can infer the presence of topological edge states and provide a spatial description of the state, revealing both the localization boundary and the penetration depth into the bulk of the edge state. The  LDOS retrieved from the scanning probe is scaled and compared to the amplitude profiles of corresponding edge states in an open-boundary isolated SSH chain in Fig.~\ref{profileES}. As expected from our interpretation of $(\tau^{nn}_\phi)^{-1}$, the decoherence rate has the same decay constant as a function of $n$ as the edge states' density profiles. These combined capabilities establish the proposed decoherence probe as a powerful tool to spatially characterize and classify localized states.
\section{Discussion and Conclusion}\label{sec:conclusions}
We have demonstrated that the decoherence of a qubit coupled to an open SSH chain environment can be used to probe states within the SSH chain subspace. This ability stems from the proportionality between the decoherence rate of the qubit with the LDOS within the SSH chain of the composite system. In particular, sweeping over the topological material of interest with a qubit probe of appropriately tuned energy, we demonstrate the ability to spatially characterize topological edge states of the SSH model through decoherence dynamics. The decoherence probe is sensitive to the presence (or absence) of edge states and determines the localization boundary of edge states as well as their spatial profile. It should be highlighted that these ideas are applicable to a wide array of systems and the results are not exclusive to the composite system studied here: the decoherence expression of the qubit    can be adapted to various environments by considering the appropriate self-energy, and will have the same form as (\ref{tauinv}) so long as a single environment-qubit coupling is considered. As such, other one-dimensional systems featuring localized states, such as the Kitaev chain hosting Majorana zero-modes \cite{Kitaev_2001}, could be spatially characterized by means of the decoherence probe. Additionally, one could envision a generalization of the probe discussed here to two dimensions. Two-dimensional topological materials, which feature edge modes along their perimeter, could be scanned over their area to study the penetration depth of edge modes as well as their location along the boundary of a sample. Corner modes, higher-order topological excitations in 2-d materials confined not to the full edges but only the corners \cite{PhysRevB.100.205406,PhysRevLett.122.076801}, should also lead to distinct signatures from the decoherence probe. It is hoped that this novel quantum sensor can become a practical platform for  studying real-space properties of localized states in low-dimensional materials and metamaterials.

\section*{Acknowledgements} \label{sec:acknowledgements}
This work was supported in part by the Natural Science and Engineering Research Council of Canada and
by the Fonds de Recherche Nature et Technologies du
Québec via the INTRIQ strategic cluster grant.

\appendix
\section{} \label{sec:appendix}
\subsection{Effective description for coupled systems}\label{appendix1}

We show here that an environment can be replaced by an effective self-energy, deriving an expression equivalent to (\ref{greensDD}) in a more general setting. Consider a system comprised of a finite subsystem coupled to a large (potentially infinite) environment/reservoir. For definiteness, consider a qubit coupled to an open channel  acting as an infinite reservoir via a single coupling $t_c$ between the second dot $|\epsilon_2\rangle$ and an arbitrary site $n$ of the lead. These results can be easily generalized to various subsystem-environment pairings afterwards. The Hamiltonian for the full system is \begin{equation}\label{Htot}
H = \left(\begin{array}{c|c}H_{\mathrm{DD}} & W \\ \hline W^{\dagger} & H_{\mathrm{R}} \end{array}\right),
\end{equation} 
where $H_{\mathrm{R}}$ is the reservoir Hamiltonian and $W$ is a 2$\times \infty$ matrix with W$^{(2,n)}=t_c$ as the only nonzero entry; the coupling between the qubit and the environment. The objective is to show that we can rigorously express
\begin{equation}\label{Gtotgoal}
G=(E-H)^{-1} = \left(\begin{array}{c|c}G_{\mathrm{DD}}' & \cdots \\ \hline \cdots & \cdots\end{array}\right),
\end{equation}
where $G_{\mathrm{DD}}' = (E-H'_{\mathrm{DD}})^{-1}$ is the $2\times2$ effective Green's function of the qubit incorporating the influence of the environment with the associated effective Hamiltonian
\begin{equation}\label{Hprimgoal}
H'_{\mathrm{DD}}=\left(\begin{array}{cc} \epsilon_1 & \tau \\ \tau & \epsilon_2 +\Sigma_{\mathrm{R}}^{nn}   \end{array}\right),
\end{equation}
where $\Sigma_{\mathrm{R}}^{nn}$ is the $n^{\mathrm{th}}$ diagonal element of the environment's self-energy matrix.

Let us separate the Hamiltonian in the following way:
\begin{equation}\label{SplitH}
H = \underbrace{\left(\begin{array}{c|c}H_{\mathrm{DD}} & 0 \\ \hline  0 & H_{\mathrm{R}} \end{array}\right)}_{H_0} + \underbrace{\left(\begin{array}{c|c}0 & W \\ \hline  W^{\dagger} & 0 \end{array}\right)}_{V}.
\end{equation} 
The block diagonal Hamiltonian matrix, $H_0$, has the corresponding Green's function
\begin{equation}\label{G0}
G_0 = (E-H_0)^{-1}=\left(\begin{array}{c|c}G_{\mathrm{DD}} & 0 \\ \hline 0 & G_{\mathrm{R}} \end{array}\right).
\end{equation}
Using this information in $G=(E-H)^{-1}$;
\begin{equation}\label{preseries}
G=(E-H_0 -V)^{-1} = (G_0^{-1}-V)^{-1} .
\end{equation}
Factoring the above into a geometric series expansion:
\begin{align}
G&=(G_0^{-1}(1-G_0 V))^{-1}=(1-G_0 V)^{-1}G_0 \nonumber\\
&= (1+G_0 V + (G_0 V)^2 + ...)G_0.
\end{align}
Note that $G_{\mathrm{DD}}'$ is a  diagonal block element of $G$ and only even powers of $G_0 V$ contribute to the diagonal block $-$ $G_0V$ being purely antidiagonal and ($G_0V)^2$ being purely diagonal. The first power of ($G_0V)^2$ gives the following
\begin{equation}
(G_0 V)^2=\left(\begin{array}{c|c}G_{\mathrm{DD}} W G_{\mathrm{R}} W^{\dagger} & 0 \\ \hline 0 & G_{\mathrm{R}}W^{\dagger} G_{\mathrm{DD}} W \end{array}\right) ,
\end{equation}
with the upper diagonal block attributed to $G_{\mathrm{DD}}'$ in $G$. Expressing $G_{\mathrm{DD}}'$ in the even powers of the series expansion then yields
\begin{align}
G_{\mathrm{DD}}'&=(1 + G_{\mathrm{DD}} W G_{\mathrm{R}} W^\dagger \nonumber\\
& \hspace{0.5cm}+(G_{\mathrm{DD}} WG_{\mathrm{R}}W^\dagger )^2 + ...)G_{\mathrm{DD}} \nonumber\\
&=(1 - (G_{\mathrm{DD}} W G_{\mathrm{R}} W^\dagger ))^{-1} G_{\mathrm{DD}} \nonumber\\
&=  (G_{\mathrm{DD}}^{-1}-G_{\mathrm{DD}}^{-1}G_{\mathrm{DD}} W G_{\mathrm{R}} W^\dagger)^{-1} \nonumber\\
&= (G_{\mathrm{DD}}^{-1} - W G_{\mathrm{R}} W^\dagger)^{-1}.
\end{align}
The product $W G_{\mathrm{R}} W^\dagger$ gives a 2$\times$2 subset of the reservoir's Green's function selecting the $n^{\mathrm{th}}$ diagonal entry, $G^{nn}_{\mathrm{R}}$:
\begin{equation}\label{WGWprod}
W G_{\mathrm{R}} W^\dagger = t^2_c G^{nn}_{\mathrm{R}} \left(\begin{array}{cc} 0 & 0 \\ 0 & 1  \end{array}\right).
\end{equation}
This allows us to express
\begin{align}
G_{\mathrm{DD}}' &= \Bigg(G_{\mathrm{DD}}^{-1}-t^2_c G^{nn}_{\mathrm{R}}\left(\begin{array}{cc} 0 & 0 \\ 0 & 1  \end{array}\right)\Bigg)^{-1} \nonumber\\
&= \Bigg(E-H_{\mathrm{DD}} -t^2_c G^{nn}_{\mathrm{R}}\left(\begin{array}{cc} 0 & 0 \\ 0 & 1  \end{array}\right)\Bigg)^{-1}.
\end{align}
We have succeeded in deriving an effective form in agreement with (\ref{Hprimgoal}) where 
\begin{equation}\label{HprimDD}
H'_{\mathrm{DD}}= H_{\mathrm{DD}} + t^2_c G^{nn}_{\mathrm{R}}\left(\begin{array}{cc} 0 & 0 \\ 0 & 1  \end{array}\right) =\left(\begin{array}{cc} \epsilon_1 & \tau \\ \tau & \epsilon_2 +\Sigma^{nn}_{\mathrm{R}}  \end{array}\right)
\end{equation}
is the effective Hamiltonian for the double dot ($\Sigma^{nn}_{\mathrm{R}} =t^2_c G^{nn}_{\mathrm{R}}$ is the reservoir's self-energy) to which we associate the effective Green's function 
\begin{equation}\label{appendixGeff}
G_{\mathrm{DD}}'=\left(\begin{array}{cc} E-\epsilon_1 & -\tau \\ -\tau & E-\epsilon_2 -\Sigma^{nn}_{\mathrm{R}}  \end{array}\right)^{-1}.
\end{equation}
As expected, this Green's function has the same form as (\ref{greensDD}). This concludes the demonstration of the effective description of a subsystem coupled to an environment.

\subsection{Green's functions $G_{\mathrm{SSH},\infty}^{nn}(E)$}\label{gnnsshinfderiv}

The diagonal elements of the $N\times N$ Green's functions of the SSH chain subspace, incorporating the influence of both the SSH chain and the two semi-infinite leads, are needed in the effective description of the qubit (\ref{effectivequbit}). These expressions depend on the parity of the SSH chain $N$ and the parity of the particular site being coupled to, $n$, corresponding to a coupling of the qubit with either the $A$ or $B$ sublattice of the SSH chain. As only diagonal elements $G_{\mathrm{SSH},\infty}^{nn}(E)$ are of interest, it is not necessary to solve for the entire matrix $G_{\mathrm{SSH},\infty}(E)$: a given $(n,n)$ element can be found from the $n^{th}$ column of $(E-H_{\mathrm{SSH},\infty})G_{\mathrm{SSH},\infty}=1$. We denote this vector $\boldsymbol{G}$ and our expression of the Green's function becomes
\begin{equation}\label{vectcolumnform}
    (E-H_{\mathrm{SSH},\infty})_{ij}\boldsymbol{G}^{jn}=\delta_{in}.
\end{equation}
Observe that, structurally, this equation is nearly identical to the Schrödinger equation with the exception of an inhomogeneity at the $n^{\mathrm{th}}$ site. We therefore approach this problem with the same translationally invariant ansatz as for the SSH model eigenstates detailed in \cite{ZAIMI2021127035}, applied here to the column vector $\boldsymbol{G}$ , with the addition of a boundary term at the $n^{\mathrm{th}}$ element of the column vector. The case for both odd $N$ and $n$ is detailed below:
\begin{align}\label{Gvectoddodd}
&\boldsymbol{G}\equiv \left(\begin{array}{c} {C_{+} e^{-i \varphi} + C_{-} e^{i \varphi} } \\ {\pm ( C_{+} e^{i \varphi} + C_{-} e^{-i \varphi})} \\ {C_{+} e^{-i \varphi}e^{i2k} + C_{-} e^{i \varphi}e^{-i2k}} \\ {\pm ( C_{+} e^{i \varphi}e^{i2k} + C_{-} e^{-i \varphi}e^{-i2k})}\\ {\vdots} \\ {C_{+} e^{-i \varphi}e^{i(2m-2)k} + C_{-} e^{i \varphi}e^{-i(2m-2)k}} \\ {\pm ( C_{+} e^{i \varphi}e^{i(2m-2)k} + C_{-} e^{-i \varphi}e^{-i(2m-2)k})} \\ {2 A_0 \rightarrow \mbox{$n^{\mathrm{th}}$ site}} \\ {\pm ( D_{+} e^{i \varphi}e^{i2mk} + D_{-} e^{-i \varphi}e^{-i2mk})} \\ {D_{+} e^{-i \varphi}e^{i(2m+2)k} + D_{-} e^{i \varphi}e^{-i(2m+2)k}} \\ {\vdots} \\ {( D_{+} e^{-i \varphi}e^{i(N-3)k} + D_{-} e^{i \varphi}e^{-i(N-3)k})}\\{\pm ( D_{+} e^{i \varphi}e^{i(N-3)k} + D_{-} e^{-i \varphi}e^{-i(N-3)k})} \\ {( D_{+} e^{-i \varphi}e^{i(N-1)k} + D_{-} e^{i \varphi}e^{-i(N-1)k})} \end{array}\right).
\end{align} Considering $n$ odd implies that $n$ corresponds to sites of the $A$ sublattice and so the boundary term in $\boldsymbol{G}$ is labelled $A_0$. We allow for the most general solution considering $k\rightarrow-k$ and $\varphi\rightarrow-\varphi$. Note here that $n$ is the site index and $m$ is the unit cell index of the SSH chain. When $n$ is odd, $n=2m+1$ and for $n$ even $n=2m+2$. The constants $C_\pm$ and $D_\pm$ are determined by boundary equations. The boundary equations of the first region (site $n=1$ to $n-1$) described by $C_\pm$ yield
\begin{equation}
\begin{gathered}
(E-\Sigma_L) G^{1n} \mp t_1 G^{2n}=0\\
\pm EG^{n-2,n}
-t_1G^{n-1,n}=t_2 G^{nn} = 2 t_2 A_0.
\end{gathered}
\end{equation}
Solving for $C_\pm$ and using the shorthand notation $\sin{(Ak)}=s_A$ yields
\begin{align}
\left(\begin{array}{l}{C_+} \\ {C_-}\end{array}\right)&= C\left(\begin{array}{l}{\mp t_2 e^{ -i \varphi} e^{i2k}} + \Sigma_L e^{i \varphi} \\ { \pm t_2 e^{i \varphi}e^{-i2k} - \Sigma_L e^{-i \varphi}}\end{array}\right), \nonumber\\
C&=\frac{E A_0}{i( \Sigma_L E s_{2m} - t_2 (t_1 s_{2m+2} +t_2 s_{2m}))}.
\end{align}
Solving for $D_\pm$ is done identically but with the second set of boundary conditions:
\begin{equation}
\begin{gathered}
\pm EG^{n+1,n}
-t_2 G^{n+2,n}=t_1 G^{nn }=2 t_1 A_0\\
(E-\Sigma_R)G^{Nn} \mp t_2 G^{N-1,n}=0,
\end{gathered}
\end{equation}
such that
\begin{align}
&\left(\begin{array}{l}{D_+} \\ {D_-}\end{array}\right)= D\left(\begin{array}{l}{ (\mp t_1 e^{ -i \varphi}}+ \Sigma_R e^{i \varphi} )e^{-i(N-1)k}\\ {(\pm t_1 e^{ i \varphi}} -\Sigma_R e^{-i \varphi} )e^{i(N-1)k}\end{array}\right), \nonumber\\
D&=\frac{E A_0}{i(t_1 (t_1 s_{N-2m-1} +t_2 s_{N-2m+1}) - \Sigma_R E s_{N-2m-1})}.
\end{align}
With both $C_\pm$ and $D_\pm$ known in terms of $A_0$, the $n^{\mathrm{th}}$ component of equation (\ref{vectcolumnform}) can be used to solve for $G^{nn}=2A_0$:
\begin{equation}
    \mp t_2 G^{n-1,n} + E G^{nn} \mp t_1 G^{n+1,n} = 1.
\end{equation}
This approach, using a column vector $\boldsymbol{G}$, finding the constants $C_\pm$ and $D_\pm$, and using this information in the $n^{th}$ equation of (\ref{vectcolumnform}) to find $G^{nn}_{\mathrm{SSH},\infty}(E)$, is identical for the three other configurations. 

For $N$ odd and $n$ odd
\begin{equation}\label{Goddodd}
\begin{split}
G^{nn}_{\mathrm{SSH},\infty}= \bigg[ E + \frac{t_{2}^{2} E s_{n-1}-t_{2} \Sigma_{L}\left(t_{1} s_{n-3}+t_{2} s_{n-1}\right)}{E \Sigma_{L} s_{n-1}-t_{2}\left(t_{1} s_{n+1}+t_{2} s_{n-1}\right)} \\ + \frac{t_{1}^{2} E s_{N-n}-t_{1} \Sigma_{R}\left(t_{1} s_{N-n}+t_{2} s_{N-n-2}\right)}{E \Sigma_{R} s_{N-n}-t_{1}\left(t_{1} s_{N-n}+t_{2} s_{N-n+2}\right)}  \bigg] ^{-1}.
\end{split}
\end{equation}
For $N$ odd and $n$ even
\begin{equation}\label{Goddeven}
\begin{split}
G^{nn}_{\mathrm{SSH},\infty}= \bigg[ E + \frac{t_{1} \Sigma_L E s_{n-2}-t_1 t_{2} \left(t_{1} s_{n}+t_{2} s_{n-2}\right)}{E t_2 s_{n}-\Sigma_L \left(t_{1} s_{n-2}+t_{2} s_{n}\right)} \\ + \frac{t_{2} E \Sigma_R s_{N-n-1}-t_{1} t_2 \left(t_{1} s_{N-n-1}+t_{2} s_{N-n+1}\right)}{E t_1 s_{N-n+1}-\Sigma_R \left(t_{1} s_{N-n+1}+t_{2} s_{N-n-1}\right)}  \bigg] ^{-1}.
\end{split}
\end{equation}
For $N$ even and $n$ odd
\begin{equation}\label{Gevenodd}
\begin{split}
G^{nn}_{\mathrm{SSH},\infty} = \bigg[ E + \frac{t_{2}^{2} E s_{n-1}-t_{2} \Sigma_{L}\left(t_{1} s_{n-3}+t_{2} s_{n-1}\right)}{E \Sigma_{L} s_{n-1}-t_{2}\left(t_{1} s_{n+1}+t_{2} s_{n-1}\right)} \\ + \frac{t_{1} E \Sigma_R s_{N-n-1 }-t_{1} t_2 \left(t_{1} s_{N-n+1}+t_{2} s_{N-n-1}\right)}{E t_2 s_{N-n+1}-\Sigma_R \left(t_{1} s_{N-n-1}+t_{2} s_{N-n+1}\right)}  \bigg] ^{-1}.
\end{split}
\end{equation}
For $N$ even and $n$ even
\begin{equation}\label{Geveneven}
\begin{split}
G^{nn}_{\mathrm{SSH},\infty}= \bigg[ E + \frac{t_{1} \Sigma_L E s_{n-2}-t_1 t_{2} \left(t_{1} s_{n}+t_{2} s_{n-2}\right)}{E t_2 s_{n}-\Sigma_L \left(t_{1} s_{n-2}+t_{2} s_{n}\right)} \\ + \frac{t_{2}^2 E  s_{N-n}-t_{2} \Sigma_R \left(t_{1} s_{N-n-2}+t_{2} s_{N-n}\right)}{E \Sigma_R s_{N-n}- t_2 \left(t_{1} s_{N-n+2}+t_{2} s_{N-n}\right)}  \bigg] ^{-1}. 
\end{split}
\end{equation}
These expressions are arranged such that the second terms stem from the interface between the left lead and the SSH chain, while the third terms correspond to the right lead/chain-interface. Note there is some symmetry in these terms; the left boundary terms for both odd-$n$ and both even-$n$ expressions are identical, which is expected as the left boundary conditions are unchanged whether $N$ is even or odd. Looking at the right boundary terms, the $N$ odd and $n$ odd term has an identical form to the $N$ even and $n$ even term but with $t_1 \leftrightarrow t_2$ due to the SSH lattice terminating on a different hopping term for even or odd chain lengths. Equivalently, notice the $N$ odd and $n$ even right boundary term also has an identical form to the $N$ even and $n$ odd case with $t_1 \leftrightarrow t_2$.

\end{document}